# Expanding possibilities of quantum state engineering and amplifying optical Schrödinger kitten state

Evgeny V. Mikheev[1], Sergey A. Podoshvedov[2*], Nguyen Ba An[3,4**]

[1]*Department of Nanoscale Physics, Institute of Natural and Exact Sciences, South Ural State University (SUSU), Lenin Av. 76, Chelyabinsk, Russia*
[2]*Laboratory of Quantum Information Processing and Quantum Computing, Institute of Natural and Exact Sciences, South Ural State University (SUSU), Lenin Av. 76, Chelyabinsk, Russia*
[3]*Thang Long Institute of Mathematics and Applied Sciences (TIMAS), Thang Long University, Nghiem Xuan Yem, Hoang Mai, Hanoi, Vietnam*
[4]*Institute of Physics, Vietnam Academy of Science and Technology (VAST), 18 Hoang Quoc Viet, Cau Giay, Hanoi, Vietnam*
[*] sapodo68@gmail.com    [**] nban@iop.vast.ac.vn

**Abstract:** We demonstrate an optical method to engineer optical Schrödinger cat states (SCSs) of large size $|\beta|$ ranging from $|\beta|=2$ to $|\beta|=3$ with high fidelity close to 0.99. Our approach uses the $\alpha$-representation of the SCSs in infinite Hilbert space with base in terms of displaced number states characterized by the displacement amplitude $\alpha$. An arbitrary $\alpha$-representation of SCSs enables manipulation of the amplitudes in wider ranges of parameters, greatly expanding the possibilities for generation of desired nonclassical states. The optical scheme we consider is quite universal for implementation of the conditioned states close to SCSs with use of linear-optics elements and detectors projecting unitarily transformed input states onto the target one. Different states (e.g., number state or coherent state or superposed state) are selected as the input to the optical scheme. In particular, an input small-size Schrödinger kitten state can give rise to an output large-size SCS.

**OCIS codes:** (270.0270) Quantum Optics; (270.5585) Quantum information and processing.

## References and links

## 1. Introduction

Quantum information processing (QIP) has dramatically changed the way we manipulate information from the world of mathematical abstract bits to the world of physical realistic qubits [1] realizable in various physical systems. Information and physical systems used for its transformation and transmission become interconnected. In order to utilize all the powerful possibilities offered by quantum mechanics for QIP, one needs to be able to effectively manipulate quantum states on physical level taking into account the peculiarities of the physical medium. This is usually a daunting task as any nontrivial manipulation with the state may be accompanied by serious limitations (possibly fundamental) that the physical system may impose. For example, optical qubits can be easily controlled, but it is difficult to get them to interact with each other in a deterministic manner [2]. The task of performing a universal quantum operation is transferred to the deterministic implementation of specific quantum state in ideal case [3]. In the case of discrete variable (DV) states, the QIP implementation is probabilistic [4-6] and, moreover, requires the initial creation of a quantum channel that can be produced off-line [7].

In continuous variables (CV) QIP, the currently available Gaussian operations [8] enable to readily manipulate any Gaussian state but the quantum protocols with Gaussian quantum channel suffer restricted fidelities of the output states [9]. In order to move out, some restricted subset such as non-Gaussian operations must be used [10-13]. Deterministic (or near-deterministic) realization of non-Gaussian operations encounters tremendous difficulties [14]. So, the mechanism of interaction between DV and CV states (DV-CV interaction mechanism) can be used to increase the efficiency of the quantum protocols [15-17]. In addition to difficulties with non-Gaussian operations, the implementation of the necessary nonclassical states different from the Gaussian ones remains problematic. Photon generation can be physically implemented with help of on-off detection that conditionally induces generation of the state with strong non-Gaussian features [18,19]. Generating more complex states, including superposed ones, is probabilistic and requires a lot of efforts, but can be performed in advance and then stored until it is needed. Currently, photon addition/subtraction techniques are commonly recognized for producing, for example, SCSs, which play an important role for quantum communication and quantum computation [20-30].

When a series of consistent photon additions (or subtractions) accompanied by the suitable displacements is used, an arbitrary superposition of the number states involving SCSs can be generated [21]. But such a technique can hardly be useful in everyday practice due to the need of a large number of optical elements and single photons being spent in order to realize SCSs of sufficiently large amplitude. Here, we develop quantum state engineering of SCSs of large size with $|\beta| \geq 2$, where β is an amplitude of quantum superpositions of two out-of-phase light pulses. The $\alpha$-representation of the SCSs is the basis for the optical technique. We can work with a set of amplitudes of SCSs in $\alpha$-representation as with amplitudes in the standard 0-representation which is formally the $\alpha$-representation when $\alpha = 0$. This interpretation of the SCSs significantly expands the possibilities of manipulating amplitudes of the nonclassical states. This allows us to more effectively engineer SCSs of large amplitude regardless of its parity with use of number states (not only single photons) in some auxiliary modes. Moreover, this approach can, to some extent, be considered universal for input states. The input states can be chosen over a wide range in order to generate SCSs of large amplitude $|\beta| \geq 2$ with high fidelity closely approaching to maximally possible upper bound. In particular, we shall show our results of the approach using different input states in the main mode, including Fock states, coherent state and Schrödinger kitten state (i.e., SCSs with amplitude as small as $|\beta| < 1$). In the case when a Schrödinger kitten is fed into the scheme as an input and the resulting output becomes a big SCS with $|\beta| \geq 2$, we can speak of its amplification [31,32] by means of photons in auxiliary modes conditioned on observing a known measurement outcome. We prove that such a gain on amplitude of an initial optical kitten is possible with high fidelity that can be regarded as an effective means of manipulating SCSs offline.

## 2. Schrödinger cat qudits

Let us consider the even/odd SCSs $|\beta_{\pm}\rangle$ with amplitude $\beta$ in $i\alpha$-representation of the even/odd SCSs determined in infinite Hilbert space of the displaced number states $\{|k, i\alpha\rangle = D(i\alpha)|k\rangle;\ k = 0,1,\dots,\infty\}$, where $\alpha$ is assumed real for simplicity, $|k\rangle$ is Fock state containing $k$ photons, while $D(\alpha) = exp(\alpha a^+ - \alpha^* a)$ with complex $\alpha$ is the displacement operator and $a^+$ ($a$) the photon creation (annihilation) operator [33]:

$$|\beta_+\rangle = N_+(|-\beta\rangle + |\beta\rangle) = N_+ exp\left(-\frac{\mathbb{a}^2}{2}\right)\sum_{k=0}^{\infty} a_k^{(+)}|k, i\alpha\rangle, \qquad (1)$$

$$|\beta_-\rangle = N_-(|-\beta\rangle - |\beta\rangle) = N_- exp\left(-\frac{\mathbb{a}^2}{2}\right)\sum_{k=0}^{\infty} a_k^{(-)}|k, i\alpha\rangle, \qquad (2)$$

with $N_{\pm} = \left(2(1 \pm exp(-2|\beta|^2))\right)^{-1/2}$ being the normalization factors, which in general depend on both $\alpha$ and $\beta$, $\mathbb{a} = \sqrt{|\alpha|^2 + |\beta|^2}$. The notations $|\pm\beta\rangle$ mean coherent states with amplitudes $\pm\beta$. The amplitudes $a_k^{(\pm)}$ of the even/odd SCS in the $i\alpha$-representation can be calculated as [34]

$$a_k^{(+)} = \frac{2(i\mathbb{a})^k}{\sqrt{k!}} cos(\alpha\beta + k(\varphi + \pi/2)), \tag{3}$$

$$a_k^{(-)} = i\frac{2(i\mathbb{a})^k}{\sqrt{k!}} sin(\alpha\beta + k(\varphi + \pi/2)), \tag{4}$$

where the relative phase is $\varphi = arctang(\alpha/\beta)$. The normalization condition $N_\pm^2 exp(-\mathbb{a}^2)\sum_{k=0}^\infty |a_k^{(\pm)}|^2 = 1$ holds for even and odd SCSs, respectively, for any values of the parameters $\alpha$ and $\beta$. The even/odd SCSs are obviously orthogonal to each other, $\langle\beta_-|\beta_+\rangle = 0$, as the photon numbers in $|\beta_+\rangle$ ($|\beta_-\rangle$) are even (odd) in the 0-representation.

Decompositions (1, 2) allow us to introduce the concept of Schrödinger cat qudits (SCQs) $|\Psi_n^{(\pm)}\rangle$ which live in an $(n + 1)$-dimensional Hilbert space with base elements $|k, i\alpha\rangle$ displaced by the quantity $i\alpha$ on phase plane with regard of the number states [34]

$$|\Psi_n^{(+)}\rangle = N_n^{(+)} \sum_{k=0}^n ((i\mathbb{a})^k/\sqrt{k!}) cos(\alpha\beta + k(\varphi + \pi/2))|k, i\alpha\rangle, \tag{5}$$

$$|\Psi_n^{(-)}\rangle = N_n^{(-)} \sum_{k=0}^n ((i\mathbb{a})^k/\sqrt{k!}) sin(\alpha\beta + k(\varphi + \pi/2))|k, i\alpha\rangle, \tag{6}$$

where $N_n^{(+)} = \left(\sum_{k=0}^n (\mathbb{a}^{2k}/k!) cos^2(\alpha\beta + k(\varphi + \pi/2))\right)^{-1/2}$ and $N_n^{(-)} = \left(\sum_{k=0}^n (\mathbb{a}^{2k}/k!) sin^2(\alpha\beta + k(\varphi + \pi/2))\right)^{-1/2}$ are the normalization factors of the SCQs. These qudits can be considered as the best approximations of genuine SCSs. The fidelity of the qudit will only increase with the increase in the number of the terms in output superposition. It should be noted that the state of the SCQs can be rewritten in the form of a product of operators $(a^+ - z_j^{(\pm SCQ)})$ as

$$|\Psi_n^{(+)}\rangle = \frac{N_n^{(+)}(i\mathbb{a})^n cos(\alpha\beta+n(\varphi+\pi/2))}{n!} D(i\alpha)\left[\prod_{j=0}^n (a^+ - z_j^{(+SCQ)})\right]|0\rangle, \tag{7}$$

$$|\Psi_n^{(-)}\rangle = \frac{N_n^{(-)}(i\mathbb{a})^n sin(\alpha\beta+n(\varphi+\pi/2))}{n!} D(i\alpha)\left[\prod_{j=0}^n (a^+ - z_j^{(-SCQ)})\right]|0\rangle, \tag{8}$$

where $z_j^{(\pm SCQ)}$ are the roots of the following polynomial

$$f_n^{(+SCQ)}(z) = \sum_{k=0}^n \frac{n!(i\mathbb{a})^{k-n}}{k!} \frac{cos(\alpha\beta+k(\varphi+\pi/2))}{cos(\alpha\beta+n(\varphi+\pi/2))} z^k, \tag{9}$$

$$f_n^{(-SCQ)}(z) = \sum_{k=0}^n \frac{n!(i\mathbb{a})^{k-n}}{k!} \frac{sin(\alpha\beta+k(\varphi+\pi/2))}{sin(\alpha\beta+n(\varphi+\pi/2))} z^k. \tag{10}$$

These polynomials can be obtained by formally replacing the creation operator $a^+$ in states (5, 6) by the variable $z$ (i.e., $a^+ \to z$). The representation of the SCQs (7, 8) are useful since the operator $(a^+ - z_j^{(\pm SCQ)})$ can be implemented using the displacement operators as $D(z_j^{(\pm SCQ)*})a^+ D^+(z_j^{(\pm SCQ)*}) = a^+ - z_j^{(\pm SCQ)}$ [33], where $D^+(z_j^{(\pm SCQ)*})$ denotes Hermitian conjugate operation to $D(z_j^{(\pm SCQ)*})$. Then, the consistent application of displacement operators with displacement amplitude equal to the root of the polynomials (9, 10) allows us to realize even/odd SCQs that can approximate SCSs of large amplitude $|\beta|$ with high fidelity

$$|\Psi_n^{(+)}\rangle = \frac{N_n^{(+)}(i\mathbb{a})^n cos(\alpha\beta+n(\varphi+\pi/2))}{n!} D(i\alpha)\left[\prod_{m=0}^n D(z_j^{(+SCQ)*})a^+ D^+(z_j^{(+SCQ)*})\right]|0\rangle, \tag{11}$$

$$|\Psi_n^{(-)}\rangle = \frac{N_n^{(-)}(i\mathbb{a})^n sin(\alpha\beta+n(\varphi+\pi/2))}{n!} D(i\alpha)\left[\prod_{m=0}^n D(z_j^{(-SCQ)*})a^+ D^+(z_j^{(-SCQ)*})\right]|0\rangle, \tag{12}$$

Thus, a set of roots $\{z_j^{(\pm SCQ)}\}$ can also definitely characterize even/odd SCQs that can be called root representation of the SCQs. In general, roots of polynomials of $n$ degree with known coefficients (9, 10) can be calculated by numerical methods. In the simplest cases of small values of $n$, these roots can be analytically obtained. So, if we consider the even/odd SCQs in 0-representation (in conventional terms of number states), we have $z_1^{(+SCQ)} = i\sqrt{2}\beta^{-1}\sqrt{3+\sqrt{3}}$, $z_2^{(+SCQ)} = -i\sqrt{2}\beta^{-1}\sqrt{3+\sqrt{3}}$, $z_3^{(+SCQ)} = i\sqrt{2}\beta^{-1}\sqrt{3-\sqrt{3}}$, $z_4^{(+SCQ)} = -i\sqrt{2}\beta^{-1}\sqrt{3-\sqrt{3}}$ for even SCQ with $n = 4$ and $z_1^{(-SCQ)} = 0$, $z_2^{(-SCQ)} = \beta^{-1}\sqrt{-10+i2\sqrt{5}}$, $z_3^{(-SCQ)} = -\beta^{-1}\sqrt{-10+i2\sqrt{5}}$, $z_4^{(-SCQ)} = \beta^{-1}\sqrt{-10-i2\sqrt{5}}$, $z_5^{(-SCQ)} = -\beta^{-1}\sqrt{-10 - i2\sqrt{5}}$ for odd SCQ with $n = 5$. Note that one root of odd SCQ polynomial is always zero. So, at least, even/odd SCQs $|\Psi_4^{(\pm)}\rangle$ can be realized by alternate application of coherent displacement operator and creation operator [21]. We are going to use the representations to efficiently generate SCQ with maximal number of the terms in the superposition.

### 3. General approach

In general case, consider the optical scheme in Fig. 1 which is adjustable for the generation of certain nonclassical states. The scheme is shown in black box style. It includes $m$ beam splitters (BSs) $BS_{0k}$, where the subscript $0k$, with $k = 1, 2, \ldots, m$, means that mode 0 and mode $k$ are superposed on a beam splitter, whose transmission (reflection) coefficients are $t_k$ ($r_k$). Action of the $BS_{0k}$ is described by the unitary matrix

$$BS_{0k} = \begin{bmatrix} t_k & r_k \\ -r_k^* & t_k^* \end{bmatrix}, \tag{13}$$

satisfying the normalization condition $|t_k|^2 + |r_k|^2 = 1$. The $BS_{0k}$ transforms creation operators in modes 0 and $k$ as $a_0^+ \to t_k a_0^+ + r_k a_k^+$ and $a_k^+ \to -r_k^* a_0^+ + t_k^* a_k^+$.

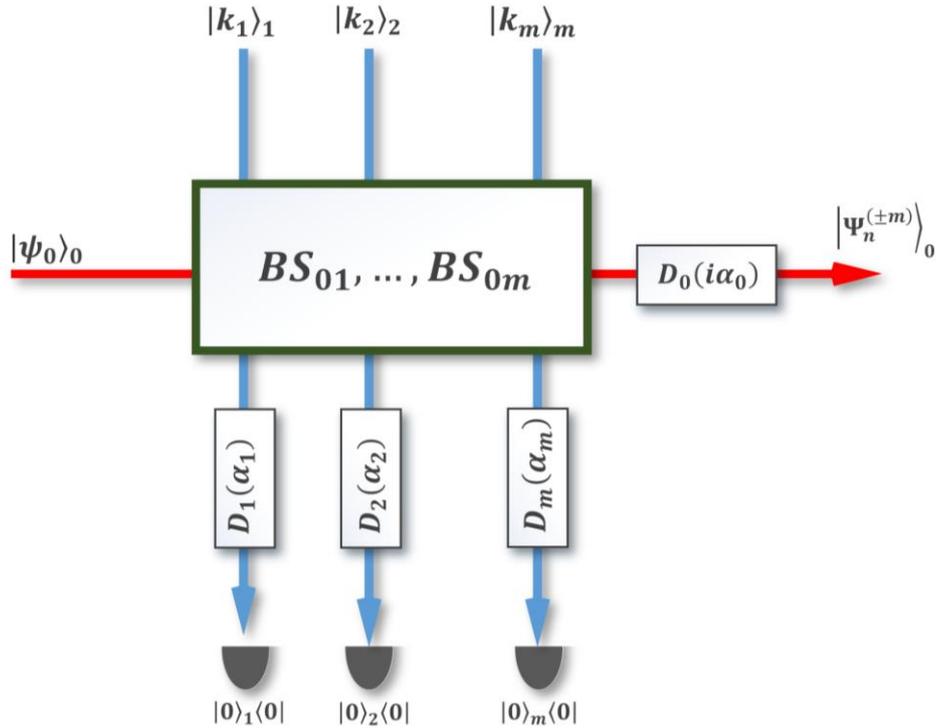

Fig. 1. Schematic setup for generation of the conditional state $|\Psi_n^{(\pm m)}\rangle$ with some probability $P_n^{(\pm m)}$ that approximate SCQs in Eqs. (6, 7) with fidelity $F_n^{(\pm m)}$ (20). The optical scheme involves black box where main mode (i.e., mode 0) is sequentially mixed with $m$ modes on beam splitters $BS_{01}, BS_{02}, \ldots, BS_{0m}$. Each mode $j$ contains $k_j$ input photons, while mode 0 may be in an arbitrary state $|\psi_0\rangle_0$ such as Fock state, coherent state or superposition state. $BS_{0k}$ means beam splitter with transmission (reflection) coefficient $t_k$ ($r_k$). The displacement operators $D_k(\alpha_k)$ and $D_0(i\alpha_0)$ with displacement amplitudes $\alpha_k$ and $i\alpha_0$ are used after post-selection giving rise to the state $|\Psi_n^{\pm m}\rangle_0$. $|0\rangle_j\langle 0|$ implies detection of no photons in mode $j$.

This black box has $m + 1$ input and $m + 1$ output modes $(0, 1, \ldots, m)$. The zeroth mode (or mode 0) is the main one where the target nonclassical state is to be generated while all remaining $m$ modes are auxiliary. $|k_1\rangle_1, |k_2\rangle_2, \ldots, |k_m\rangle_m$ are Fock states with $k_1, k_2, \ldots, k_m \geq 0$ being the number of photons in modes $1, 2, \ldots, m$, respectively. Then $n = k_1 + k_2 + \cdots + k_m$ is the total photon number inputted to the $m$ auxiliary modes. As for the state $|\psi_0\rangle_0$ inputted to mode 0, it may be Fock state, coherent state or some superposed state. The auxiliary modes are exposed to action of the displacement operators $D_1(\alpha_1), D_2(\alpha_2), \ldots, D_m(\alpha_m)$ after mixing on $m$ BSs, where $D_k(\alpha_k)$ is the displacement operator [33] with arbitrary displacement amplitudes $\alpha_1, \alpha_2, \ldots, \alpha_m$ (the role of the displacement operators will be shown below). Note also that part of the displacement operators (not all) may be used in the optical scheme in Fig. 1 depending on the goal. Let us only remark that the displacement operation can be realized by mixing the target state with a strong coherent state on a highly transmissive beam splitter (HTBS) [35-37]. Post-selection is produced by a set of $m$ detectors $\{|l_k\rangle_k\langle l_k|; k = 1, 2, \ldots, m; l_k \geq 0\}$ in all $m$ auxiliary modes to project the main output mode of the "black box" onto the desired state. We are interested in observing the event when all the $m$ detectors do not register any photons (i.e., $l_k = 0$ for all $k = 1, 2, \ldots, m$). The final displacement operator $D_0(i\alpha_0)$ with displacement amplitude $i\alpha_0$ is applied to the conditional state to make it close to SCQs in $i\alpha_0$-representation. Final displacement operator $D_0(i\alpha_0)$ may not be used in which case we can say that the generated state is similar to the SCQs with accuracy up to the displacement operator.

The full initial state can be written as

$$|\Phi_n^{(m)}\rangle = \frac{f_0(a_0^+)a_1^{+k_1}a_2^{+k_2}\ldots a_m^{+k_m}}{\sqrt{k_1!\ldots k_m!}}|00\ldots0\rangle_{01\ldots m}, \qquad (14)$$

where $f_0(a_0^+)$ is some polynomial determining the state $|\psi_0\rangle_0$ inputted to mode 0. The state before measurement becomes

$$\rho = D_m(\alpha_m)\ldots D_1(\alpha_1)BS_{0m}\ldots BS_{01}\rho_{in}(D_m(\alpha_m)\ldots D_1(\alpha_1)BS_{0m}\ldots BS_{01})^+,$$

where $\rho_{in} = |\Phi_n^{(m)}\rangle\langle\Phi_n^{(m)}|$. Finally, we can write the density matrix of the conditioned state $|\Psi_n^{(\pm m)}\rangle$ in case of absence of any clicks (i.e., registration of the vacuum state) in all the auxiliary modes as

$$\rho_{out} = D_0(i\alpha_0)\frac{Tr_{1,2,\ldots m}(\Pi_1\Pi_2\ldots\Pi_m\rho)}{Tr_{0,1,\ldots m}(\Pi_1\Pi_2\ldots\Pi_m\rho)}D_0^+(i\alpha_0), \qquad (15)$$

where $\Pi_j = |0\rangle_j\langle 0|$ is the projection onto vacuum in mode $j$. $Tr_{1,2,\ldots m}$ and $Tr_{0,1,\ldots m}$ denote trace operations over the base states in modes $1, 2, \ldots, m$ and modes $0, 1, \ldots, m$, respectively. Expression $P_n^{(\pm m)} = Tr_{0,1,\ldots m}(\Pi_1\Pi_2\ldots\Pi_m\rho)$ defines success probability to generate the conditional state.

Let us define a measure of the closeness between the generated state and target SCSs in Eqs. (1, 2) through the fidelity

$$F_n^{(\pm m)} = tr(\rho_{out}\rho_{SCS}^{(\pm)}), \qquad (16)$$

where $\rho_{SCS}^{(\pm)}$ is a density matrix of the SCSs, subscript $n$ stands for the total number of photons in all the auxiliary modes, superscript $m$ refers to number of auxiliary modes and symbols $\pm$ denote even/odd SCQ. The fidelity lies in the range from 0 up to 1. We are interested in finding such experimental conditions (beam splitter parameters of the black box and displacement amplitudes of the displacement operators) that provide the highest possible fidelities $F_n^{(\pm m)} \approx 1$ which ensures that the generated states and SCQs will slightly differ from each other.

We can reproduce the final state (15) in the form convenient for the numerical search. The initial state (14) after action of "black box" beam splitters can be rewritten as $BS_{0m}\ldots BS_{02}BS_{01}|\Phi_n^{(m)}\rangle = BS_{0m}\ldots BS_{02}BS_{01}f_0(a_0^+)f_1(a_1^+)\ldots f_m(a_m^+)|00\ldots0\rangle_{01\ldots m} = f(a_0^+, a_1^+\ldots, a_m^+)|00\ldots0\rangle_{01\ldots m}$, where $f_0, f_1, \ldots, f_m$ are the polynomials over the creation operator, for example, $f_1(a_1^+) = a_1^{+k_1}/\sqrt{k_1!}$, $f_2(a_2^+) = a_2^{+k_2}/\sqrt{k_2!}$, ..., $f_m(a_m^+) = a_m^{+k_m}/\sqrt{k_m!}$. The resulting polynomial $f(a_0^+, a_1^+\ldots, a_m^+)$ depends on both creation operators $a_i^+$ and beam splitter parameters of the black box. If we use displacement operators and consider only the state that appears as a result of the registration of the vacuum in all the auxiliary modes, then it has the following (non-normalized) form $|\Psi\rangle_0 = \langle 00\ldots0|_{12\ldots m}D_0(i\alpha_0)D_1(i\alpha_1)\ldots D_n(\alpha_n)f(a_0^+, a_1^+\ldots a_m^+)|00\ldots0\rangle_{01\ldots m}$ that is transformed to $|\Psi\rangle_0 = exp(-(|\alpha_1|^2 + \cdots + |\alpha_n|^2)/2)D_0(i\alpha_0)f(a_0^+, -\alpha_1^*\ldots -\alpha_n^*)|0\rangle_0$, by virtue of the following relations $\langle 0|D(\alpha) = \langle -\alpha|$ and $\langle \alpha|f(a^+) = f(\alpha^*)\langle \alpha|$. If we decompose the polynomial $f(a_0^+, a_1^+ ..., a_m^+)$ into a Taylor series in powers of the creation operator $a_0^+$, we obtain the output state

$$|\Psi_n^{(\pm m)}\rangle_0 = \frac{1}{\sqrt{N_n^{(\pm m)}}} \sum_{l=0}^\infty \frac{f^{(l)}(0, -\alpha_1^*-\alpha_2^*...-\alpha_n^*)}{\sqrt{l!}} |l, i\alpha_0\rangle_0, \qquad (17)$$

where $f^{(l)}$ is the $l$-th derivative of the polynomial over $a_0^+$ and $N_n^{(\pm m)}$ is the normalization factor. Then, the success probability of the interested event becomes $P_n^{(\pm m)} = exp(-(|\alpha_1|^2 + |\alpha_2|^2 + \cdots + |\alpha_n|^2))N_n^{(\pm m)}$. The form of the $l$-th derivative and, accordingly, the conditional state (17) and the success probability of its generation can be calculated numerically for an arbitrary input state $|\psi_0\rangle_0$. Using the functions $f^{(l)}(0, -\alpha_1^* - \alpha_2^* ... - \alpha_n^*)$, we can also calculate the fidelities (16).

## 4. Coherent state and number state as input

As mentioned above, the state inputted to mode 0 of the black box of the optical scheme in Fig. 1 can be arbitrary. Let it be a Fock state $|\psi_0\rangle_0 = |k_0\rangle_0$ or a coherent state $|\psi_0\rangle_0 = |\gamma\rangle_0$. Then, the generated state $|\Psi_n^{(\pm m)}\rangle$ turns out to be represented as the result of the action of the product of operators on the vacuum, namely

$$|\Psi_n^{(\pm m)}\rangle_0 = N_n^{(\pm m)} D_0(i\alpha_0) \sum_{k=0}^n b_k^{(\pm)} |k\rangle_0 =$$
$$N_n^{(\pm m)} \exp(i\phi_n^{(\pm)}) D_0(i\alpha_0) \left[\prod_{j=0}^m (a_0^+ - z_j^{(\pm m)})^{k_j}\right]|0\rangle_0, \qquad (18)$$

where $k_j$ exactly equals the number of photons in the auxiliary mode $j$, the numbers $z_j^{(\pm m)}$ depend in a complex way on the parameters of the beam splitters and displacement amplitudes, $b_k^{(\pm)}$ are the amplitudes of the superposition (18) and $\phi_n^{(\pm)}$ is the phase of the amplitude $b_k^{(\pm)} = |b_k^{(\pm)}|\exp(i\phi_n^{(\pm)})$ and $N_n^{(\pm m)}$ is normalization factor of the state. This form (18) may resemble the root representation of the SCQs, especially in the special case with $k_1 = k_2 = \cdots = k_m = 1$.

As illustrating examples, let us first derive the formula (18) when $|\psi_0\rangle_0 = |\gamma\rangle_0$, $m = 1$ and $n = k_1$. Then, we have the chain of transformations

$$D(i\alpha_0)D_1(\alpha_1)BS_{01}(|\gamma\rangle_0|k_1\rangle_1) = \frac{1}{\sqrt{k_1!}} D_0(i\alpha_0)D_0(\gamma t_1)D_1(\gamma r_1)(-r_1^* a_0^+ + t_1^* a_1^+)^{k_1}|00\rangle_{01}$$
$$= \frac{exp(i\varphi_1)}{\sqrt{k_1!}} D_0(i\alpha_0)D_0(\gamma t_1)\left(-r_1^* a_0^+ + t_1^*(a_1^+ - \alpha_1^* - \gamma^* r_1^*)\right)^{k_1}|0\rangle_0|\alpha_1 + \gamma r_1\rangle_1.$$

Projecting the obtained state on the vacuum state of the auxiliary mode $\Pi_1 = |0\rangle_1\langle 0|$, we obtain a (non-normalized) state

$$|\Delta_n^{(1)}\rangle = \frac{exp(i\varphi_1)}{\sqrt{k_1!}} exp\left(-\frac{|\alpha_1+\gamma r_1|^2}{2}\right)(-r_1^*)^{k_1} D_0(i\alpha_0)D_0(\gamma t_1)\left(a_0^+ - z_1^{(1)}\right)^{k_1}|0\rangle_0, \qquad (19)$$

where $z_1^{(1)} = -(t_1(\alpha_1 + \gamma r_1)/r_1)^*$ and $\varphi_1 = Im(\alpha_1 \gamma^* r_1^*)$. Normalizing the state (19), we obtain either the state $|\Psi_n^{(+m)}\rangle$ or $|\Psi_n^{(-m)}\rangle$ which is of the form of Eq. (18). Next, we derive the formula (18) when $|\psi_0\rangle_0 = |\gamma\rangle_0$, but $m = 2$ and $n = k_1 + k_2$. Using the same technique, we have

$$|\Delta_n^{(2)}\rangle = \frac{exp(i(\varphi_1+\varphi_2))}{\sqrt{k_1!k_2!}} exp\left(-\frac{|\alpha_1+\gamma r_1|^2}{2}\right) exp\left(-\frac{|\alpha_2+\gamma t_1 r_2|^2}{2}\right)(-r_1^* t_2)^{k_1}(-r_2^*)^{k_1}$$
$$D_0(i\alpha_0)D_0(\gamma t_1 t_2)\left(a_0^+ - z_1^{(2)}\right)^{k_1}\left(a_0^+ - z_2^{(2)}\right)^{k_2}|0\rangle_0, \qquad (20)$$

where $z_1^{(2)} = r_2(\alpha_2 + \gamma t_1 r_2)^*/t_2 - t_1^*(\alpha_1 + \gamma r_1)^*/(r_1^* t_2)$, $z_2^{(2)} = -t_2^*(\alpha_2 + \gamma t_1 r_2)^*/r_2^*$, $\varphi_1 = Im(\alpha_1 \gamma^* r_1^*)$ and $\varphi_2 = Im(\alpha_2 \gamma^* t_1^* r_2^*)$. If we introduce the norm for the state $|\Delta_n^{(2)}\rangle$, we obtain the final state which is again of the form of Eq. (18). Applying the technique, it can be verified that formula (18) takes place in the most general case of an arbitrary number $m$ of auxiliary modes and the total number of photons in them $n = k_1 + \cdots + k_m$. The formula (18) also holds when $|\psi_0\rangle_0 = |k_0\rangle_0$ in which case $n = k_0 + k_1 + \cdots + k_m$ and an extra term $\left(a_0^+ - z_0^{(\pm m)}\right)^{k_0}$ is added. It is worth noting that if we take $\gamma = 0$ (i.e., $|\psi_0\rangle_0 = |0\rangle_0$ being the vacuum as input to the main mode) $\alpha_1 = \alpha_2 = \cdots = \alpha_m = 0$, then we have $z_0^{(\pm m)} = z_1^{(\pm m)} = \cdots = z_m^{(\pm m)} = 0$, which means that we get Fock state $|n\rangle_0$ at the output instead of the superposition (18). Nevertheless, it is possible to use the condition $\alpha_1 = \alpha_2 = \cdots = \alpha_m = 0$ in the case of an input being a coherent state $|\psi_0\rangle_0 = |\gamma\rangle_0$ with $\gamma \neq 0$, since $z_j^{(m)}$ are not equal to zero. For example, the above values $z_j^{(m)}$ become $z_1^{(1)} = -(\gamma r_1)^*$, $z_1^{(2)} = -t_1^* \gamma^* |t_2|^2 / t_2$ and $z_2^{(2)} = -t_1^* t_2^* \gamma^*$. In the case of the vacuum state in the main mode $|\psi_0\rangle_0 = |0\rangle_0$, we have $z_1^{(1)} = -(t_1 \alpha_1 / r_1)^*$, $z_1^{(2)} = (r_1^* r_2 \alpha_2 - t_1^* \alpha_1^*)/(r_1^* t_2)$ and $z_1^{(2)} = t_2^* \alpha_2^* / r_2^*$.

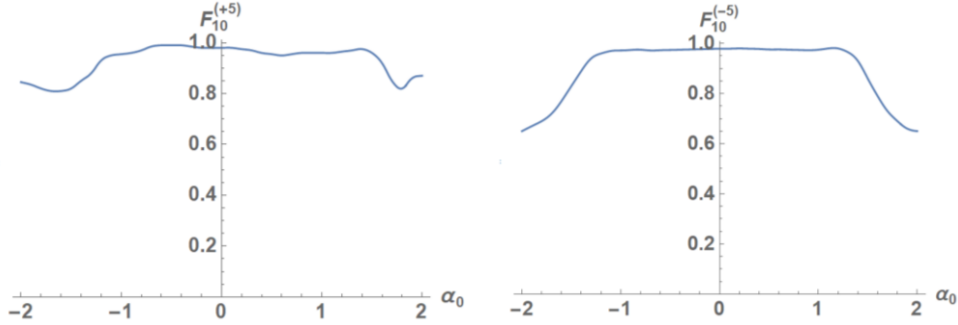

Fig. 2. Dependencies of the fidelities $F_{10}^{(+5)}$ (left) and $F_{10}^{(-5)}$ (right) between displaced qudits generated in the optical scheme in Fig. 1 and even/odd SCS of amplitude $\beta = 2$ on the displaced amplitude $\alpha_0$.

If we choose $k_1 = k_2 = \cdots = k_m = 1$ (i.e., $n = m$) in the optical scheme in Fig. 1, then it is possible to always choose experimental parameters so that $z_1^{(\pm SCQ)} = z_1^{(\pm m)}$, $z_2^{(\pm SCQ)} = z_2^{(\pm m)}$, ... and $z_m^{(\pm SCQ)} = z_m^{(\pm m)}$. Thus, if we apply additional phase factor $i^n$ to it, the conditional state (18) becomes nothing else but the SCQs as defined in Eqs. (5-8). But such a realization of the SCQs is hardly appropriate in practical experiments because of the complexity of its implementation and the need of a large number of single photons that must be spent to generate the states. To save on resources, we may assume that roots of the SCQs just slightly differ from each other, i.e., $z_j^{(\pm SCQ)} \approx z_i^{(\pm SCQ)}$ with $j \neq i$, to approximate in Eq. (18) the operator product $\left(a_0^+ - z_j^{(+SCQ)}\right)\left(a_0^+ - z_{j+1}^{(+SCQ)}\right) \ldots \left(a_0^+ - z_{j+k_j}^{(+SCQ)}\right)$ by $\left(a_0^+ - z_j^{(\pm m)}\right)^{k_j}$, where the $z_j^{(\pm m)}$ is the $k$-multiple root of some polynomial. Consider it on the example of the case with $k_0 = 0, k_1 = k_2 = k_3 = k_4 = k_5 = 2$, i.e., $m = 5$ and $n = 10$. Corresponding plots of the fidelities $F_{10}^{(\pm 5)}$ between conditional states (18) and genuine SCSs (1, 2) with amplitude $\beta = 2$ in dependence on parameter $\alpha_0$ are shown in Fig. 2. The figure indicates that there is a large range of the values $\alpha_0$ for which SCQs of large amplitude can be realized with high fidelity. Despite the fact that the fidelity can reach quite high values $F_{10}^{(+5)} > 0.98$, this implementation requires modification by reducing the number of optical elements.

In what follows, let us deal with generation of the conditional states $\left|\Psi_n^{(\pm m)}\right\rangle$ for three sets of parameters $\{m, k_0, k_1, \ldots, k_m\}$:
(i)      $m = 3, k_0 = 4, k_1 = k_2 = k_3 = 2$, (i.e., 3 BSs and $4 + 3 \times 2 = 10$ photons),
(ii)      $m = 3, k_0 = k_1 = k_2 = k_3 = 3$, (i.e., 3 BSs and $4 \times 3 = 12$ photons),
(iii)      $m = 2, k_0 = k_1 = k_2 = 4$, (i.e., 2 BSs and $3 \times 4 = 12$ photons).

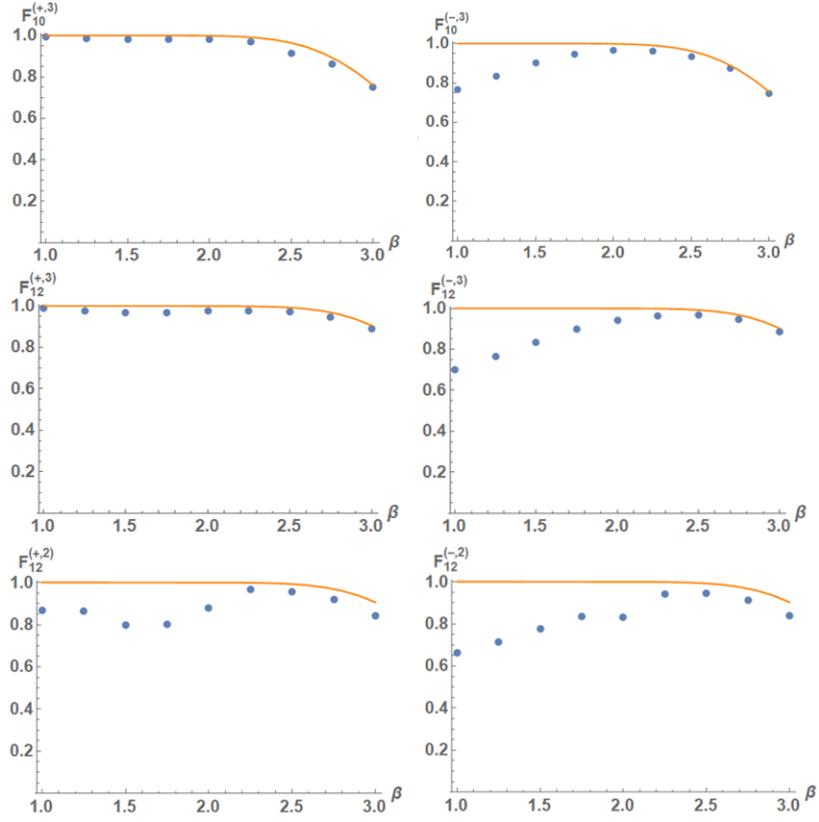

Fig. 3. Result of numerical simulation (points) of dependencies of the fidelities $F_{10}^{(\pm 3)}$ (3 BSs and 10 photons in total), $F_{12}^{(\pm 3)}$ (3 BSs and 12 photons in total) and $F_{12}^{(\pm 2)}$ (2 BSs and 12 photons in total) on $\beta$. The values obtained are upper-bound by the fidelities of genuine SCQs in Eqs. (5, 6) (solid curves).

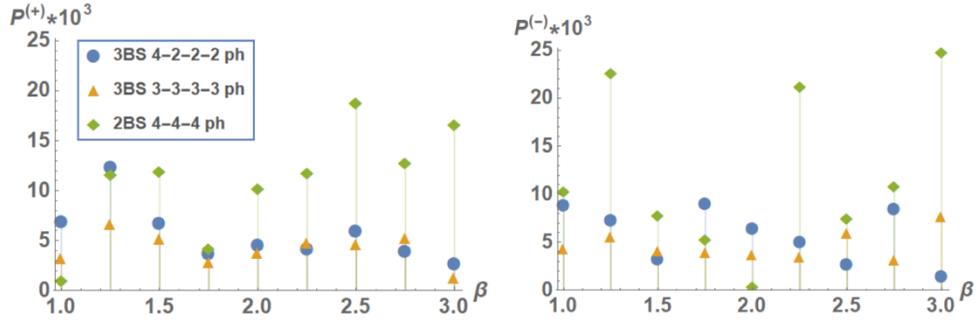

Fig. 4. Success probabilities $P_{10}^{(\pm 3)}$, $P_{12}^{(\pm 3)}$ and $P_{12}^{(\pm 2)}$ to produce displaced qudits in optical scheme in Fig. 1 that approximate SCS with fidelity shown in Fig.3.

|  | $m = 2$, $k_0 = k_1 = k_2 = 4$ | | $m = 3, k_0 = k_1 = k_2 = k_3 = 3$ | | $m = 3, k_0 = 4, k_1 = k_2 = k_3 = 2$ | |
| --- | --- | --- | --- | --- | --- | --- |
|  | $\|\beta_+\rangle$ | $\|\beta_-\rangle$ | $\|\beta_+\rangle$ | $\|\beta_-\rangle$ | $\|\beta_+\rangle$ | $\|\beta_-\rangle$ |
| $\beta$ | 2.5 | 2.5 | 2.5 | 2.5 | 2.25 | 2.25 |
| $Fidelity$ | 0.957 | 0.947 | 0.971 | 0.967 | 0.97 | 0.963 |
| $P \times 10^3$ | 18.8 | 7.5 | 4.7 | 6 | 4.2 | 5.1 |

Table 1. Numerical values of the fidelities and success probabilities $P$ of the generated displaced qudits.

Corresponding dependencies of the fidelities $F_{10}^{(\pm 3)}$, $F_{12}^{(\pm 3)}$ and $F_{12}^{(\pm 2)}$ on amplitude $\beta$ of SCSs are shown in Fig. 3 as point values. These values are upper-bound by the solid curves corresponding to the fidelity between genuine SCQs (5, 6) and SCSs (1, 2). As can be seen from the figure, the resulting numerical values of the fidelity are quite close to its upper boundary in the case of large $\beta$ values, especially for $F_{10}^{(\pm 3)}$ and $F_{12}^{(\pm 3)}$ in the cases (i) and (ii). A decrease in the number of optical elements as in the case (iii) leads to the decrease in the numerical values of the fidelity $F_{12}^{(\pm 2)}$. Despite the fact that the fidelity of the generated qudit is reduced, the optical scheme with the parameters as in the case (iii) can be viewed as promising from the point of view of the effectiveness of its implementation. Note also the fact that this scheme can only provide high fidelity values being as much as possible only for certain values of $\beta$, especially, for odd SCS. As can be seen from the right plots in Fig. 3, there are domains of $\beta$, where the fidelity between the generated state and the odd optical SCS is insufficient. In general, we can assume that this approach with a large number of photons in auxiliary modes is acceptable for the generation of SCQs of large amplitude $\beta > 2$. Let us give in Table 1 the numerical values of the parameters of generated qudits in the optical scheme in Fig. 1.

We also show (point graphs) the dependence of the success probabilities $P_{10}^{(\pm 3)}$, $P_{12}^{(\pm 3)}$ and $P_{12}^{(\pm 2)}$ on the size $\beta$ of the SCSs in Fig. 4. As can be seen from these plots, decrease in the number of the optical elements (BS, photodetectors) increases the success probability to generate large-sized SCQs. Despite low values of the success probabilities to generate the conditional states, it would not be a serious obstacle in realization of the SCQs as we can generate them off-line.

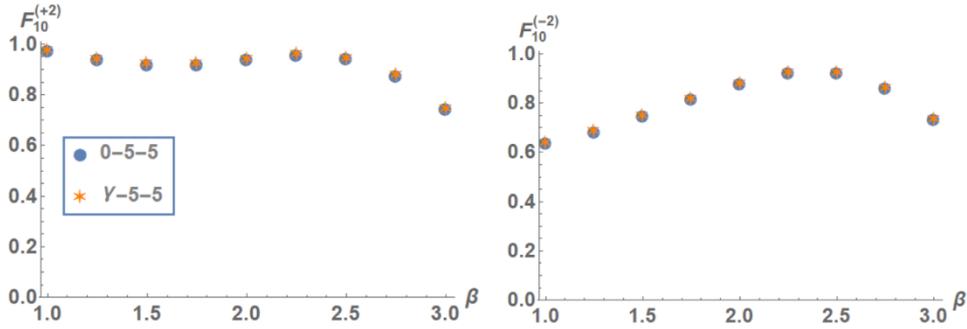

Fig. 5. Result of numerical simulation (points) of dependencies of the fidelities $F_{10}^{(\pm 2)}$ and $F_{12}^{(\pm 3)}$ on $\beta$. Schemes with the same number of photons in the auxiliary modes and with the input state either vacuum $(0 - 5 - 5)$ or coherent state $(\gamma - 5 - 5)$ are considered and compared.

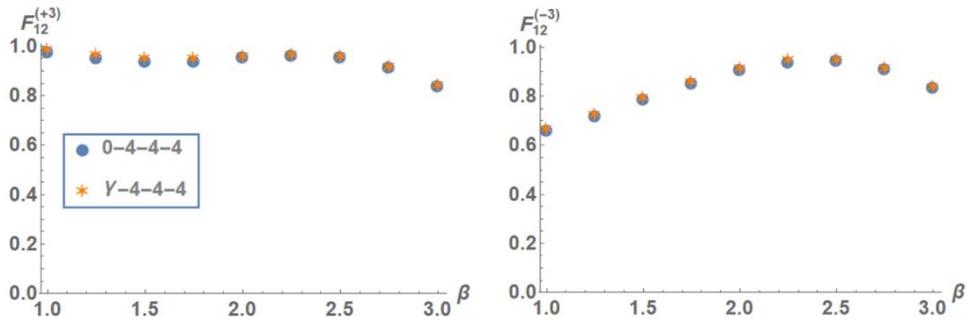

Fig. 6. Result of numerical simulation (points) of dependencies of the fidelities $F_{10}^{(\pm 2)}$ and $F_{12}^{(\pm 3)}$ on $\beta$. Schemes with the same number of photons in the auxiliary modes and with the input state either vacuum $(0 - 4 - 4 - 4)$ or coherent state $(\gamma - 4 - 4 - 4)$ are considered and compared.

We now turn to consider the implementation of the even/odd SCQs if the coherent state with certain initial amplitude is used as the input to the main mode of the "black box" in Fig. 1. Then, the conditioned output state is also representable in the form (18), where some roots $z_j^{(\pm m)}$ are analytically calculated in some simple cases (Eqs. (19, 20)). In the most general case, the parameters of the black box and displacement amplitudes that ensure the highest possible fidelity of the generated state with even/odd SCSs are sought by numerical methods. We made the appropriate mathematical modeling for some of the cases of generation of the conditional states. The corresponding fidelities $F_{10}^{(\pm 2)}$ and $F_{12}^{(\pm 3)}$ are shown in Figs. 5 and 6 for the input total states $|\varphi\rangle_{012} = |\gamma\rangle_0|5\rangle_1|5\rangle_2$ compared with $|\varphi'\rangle_{012} = |0\rangle_0|5\rangle_1|5\rangle_2$ (see Fig. 5) and $|\chi\rangle_{0123} = |\gamma\rangle_0|4\rangle_1|4\rangle_2|4\rangle_3$ compared with $|\chi'\rangle_{0123} = |0\rangle_0|4\rangle_1|4\rangle_2|4\rangle_3$ (see Fig. 6), respectively. In general, the obtained values of the fidelities, with and without using the initial coherent state almost coincide. But at certain values of the amplitude $\beta$ of the target SCS, a slight (within 0.05) increase in the fidelity of the conditioned state is observed. We gave two examples of the possible influence of the input coherent state on the output conditioned state in the case of the same states in auxiliary modes. But, generally, the influence of the initial coherent state may be enhanced by considering different photon states in auxiliary modes. From the plots in Figs. 5 and 6 we can see that this approach is applicable only to the generation of even/odd SCSs in a certain range of its size $\beta$, namely, in the case of large values $\beta > 2$. It is also worth noting that there is a deterioration in the fidelity of the conditioned state in the event of a further increase of the amplitude $\beta > 2.5$.

## 5. Schrödinger kitten state as input

In this section let us consider SCSs, Eqs. (1, 2), of small amplitude $\beta_{in} \leq 1$ (called optical Schrödinger kitten state) as input to the main mode of the black box in Fig. 1. Then the problem we are interested in can be stated as follows: whether is it possible to transform the kitten state with $\beta_{in} \leq 1$ to an adult SCS with the maximum possible amplitude $\beta \geq 2$ with highest fidelity? In this case, we can talk about amplifying size of the initial optical kitten [31, 32]. If such a transformation is possible, then how many times in amplitude it needs to amplify the initial optical kitten to an adult cat of desired large-enough size? The formulation of such a problem makes sense since an optical kitten can be very well approximated by the superposition of the first terms in Eqs. (1, 2) [22]. For such states, one can even consider the superposition of a vacuum 'plus' a two-photon state $|0\rangle + \beta^2/\sqrt{2}|2\rangle$ instead of original even kitten and a single-photon 'plus' a three-photon state $|1\rangle + \beta^2/\sqrt{3!}|3\rangle$ instead of original odd kitten, respectively. Nevertheless, in order to generate the so-called adult SCS, we believe that an optical kitten, either even or odd (see Eqs. (1, 2)), that could be produced in practice by different methods, can be used as an input state to the main mode of the black box in Fig. 1. Since the results of numerical simulations with an input coherent state give a positive result as shown in the previous section (see Fig. 4), it can be expected that positive results can be observed as well in the case of coherent superposition on the input.

In this case, the use of the formula (15) is difficult as input involves superposition state and here we cannot use the decomposition of the output state as a product of operator factors $(a_0^+ - z_j^{(\pm m)})$ as has been done in the case of input Fock or coherent states (Eqs. 18-20). Therefore, one should make use of the more specific expression (17) adjusted for numerical simulation of the problems of such kind. The results of numerical simulations for finding the global maximum of the fidelity $F^{(+2)} \equiv F_n^{(+2)}$ and $F^{(-2)} \equiv F_n^{(-2)}$, where $n$ is the total number of auxiliary photons are shown in Figs. 7 and 8 for initial optical kitten with amplitude $\beta_{in} = 0.5$ (see Fig. 7) and $\beta_{in} = 1$ (see Fig. 8), respectively. The black box in Fig. 7 is made up of two connected BSs with input pairs of photons $(4 - 4)$. The size of the initial kitten is "growing" to an acceptable size, but the fidelity of the output state in Fig. 7 may be insufficient for practical application. Therefore, it is worth considering increasing the size of the initial kitten to $\beta_{in} = 1$ as shown in Fig. 8. The black box consisting of a system of coupled BSs with input pairs of photons $(1 - 1)$, $(2 - 2)$ and $(4 - 4)$, respectively, is considered for comparison in Fig. 8. It should be noted that the results of numerical simulation of the fidelity $F^{(+2)}$ fall quite smoothly from the maximum value 1 to lower ones for all considered cases. The fidelity $F^{(-2)}$ behaves in approximately the same way as in Figs. 3, 5 and 6, except of the case $(1-1)$. It starts from values less than one, reaches its maximum and then falls down. But, nevertheless, the greatest fidelity is observed when using a pair $(4-4)$ of auxiliary photons. In general, the following picture is observed: the greater the number of auxiliary photons is taken, the larger the amplitude and the higher the fidelity of the generated SCS are obtained. Note that a smaller number of auxiliary photons is required to engineer the state in the case of an input optical kitten, unlike the case of the input Fock and coherent states. Dotted plots of the success probabilities $P^{(+2)} \equiv P_n^{(+2)}$ and $P^{(-2)} \equiv P_n^{(-2)}$, where $n$ is the total number of auxiliary photons, to implement the conditional states are shown in Figs. 9 and 10, respectively. These graphics correspond to the data presented on Figs. 7 and 8. In general, there is an increase in the success probability compared with the data in Fig. 4 and in Table 1, which is due to the fact that the lesser number of optical elements is used. We also note sufficiently large probabilities of success in the case of using an auxiliary pair of four-photon states. Tables 2 and 3 present some numerical values for $F^{(\pm 2)}$ and $P^{(\pm 2)}$ following from the plots. Analyzing the obtained results, it can be conjectured that the amplitude and fidelity of the conditional states will only increase with an increase in either the number of auxiliary photons or size of the input kitten state. As for the amplification of an optical kitten, this can only be assessed qualitatively because a quantitative measure of amplification has not yet been proposed. Nevertheless, we could take the ratio of $\beta/\beta_{in}$, when the maximum fidelity of the conditional state is observed, as a possible measure. Then it is easier to estimate it in the case of optical scheme $(4-4)$ with input optical kitten with $\beta_{in} = 1$. In this particular case the amplification measure is $\beta/\beta_{in} = 2.5$, implying that a kitten of size $\beta_{in} = 1$ is grown up to an adult cat of size $\beta = 2.5$.

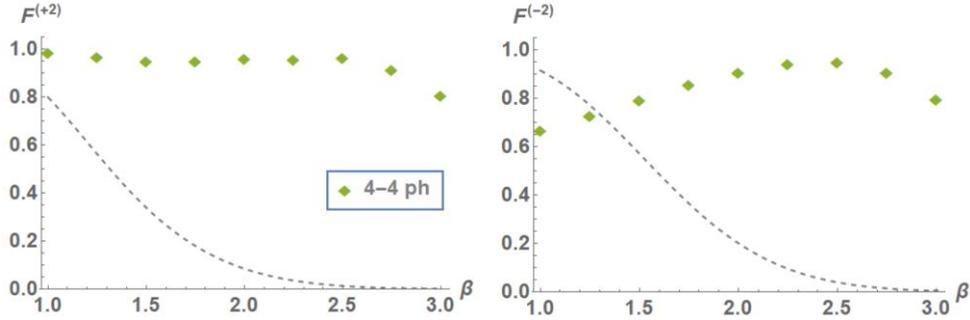

Fig. 7. Result of numerical simulation (points) of dependencies of the fidelities $F^{(+2)} \equiv F_8^{(+2)}$ and $F^{(-2)} \equiv F_8^{(-2)}$ on $\beta$. Two connected BSs with corresponding auxiliary photons are used. The original kitten with $\beta_{in} = 0.5$ is launched to the input of the system in Fig. 1. The dotted curve is the theoretical fidelity between the original kitten and SCSs with the corresponding amplitude.

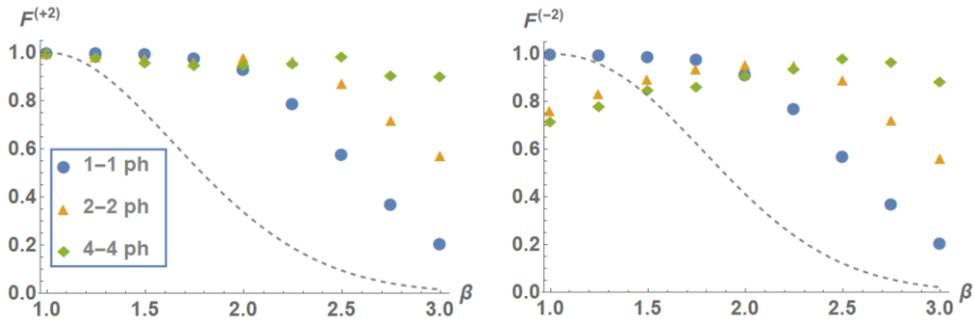

Fig. 8. Result of numerical simulation (points) of dependencies of the fidelities $F^{(+2)} \equiv F_n^{(+2)}$ and $F^{(-2)} \equiv F_n^{(-2)}$, where $n$ is total number of auxiliary photons, on $\beta$. Two connected BSs with corresponding auxiliary photons are used. The original kitten with $\beta = 1$ is launched to the input of the system in Fig. 1. The dotted curve is the theoretical fidelity between the original kitten and SCSs with the corresponding amplitude.

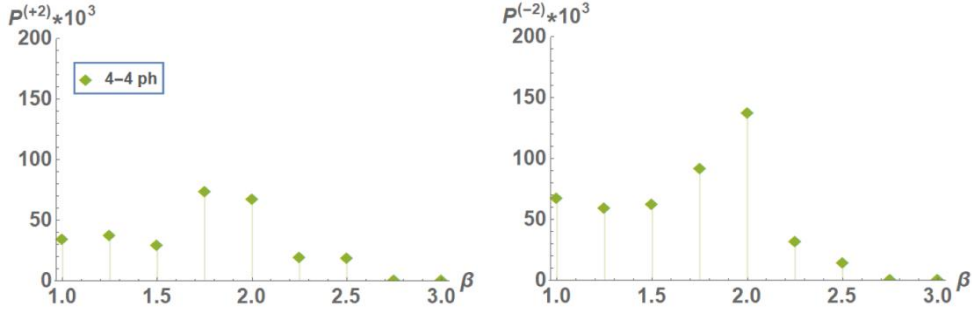

Fig. 9. Result of numerical simulation (points) of dependencies of success probabilities $P^{(+2)} \equiv P_8^{(+2)}$ and $P^{(-2)} \equiv F_8^{(-2)}$ of realization of the conditional states in Fig. 7 from optical kitten with $\beta_{in} = 0.5$ on $\beta$.

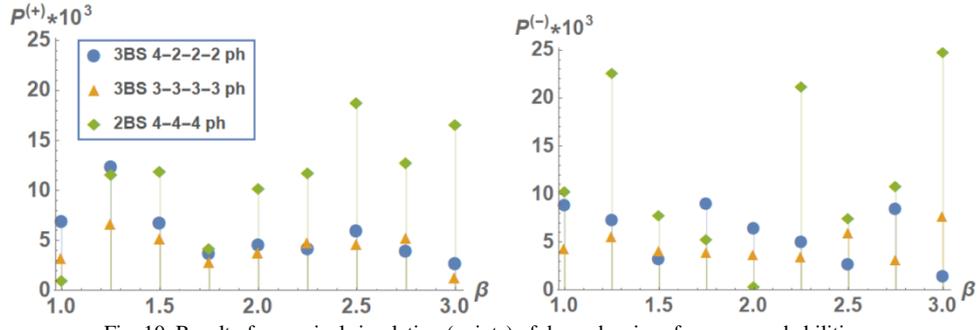

Fig. 10. Result of numerical simulation (points) of dependencies of success probabilities $P^{(+2)} \equiv P_n^{(+2)}$ and $P^{(-2)} \equiv P_n^{(-2)}$, where $n$ is total number of auxiliary photons, of realization of the conditional states in Fig. 8 from optical kitten with $\beta_{in} = 1$ on $\beta$.

|  | 4-4 | |
|---|---|---|
|  | $|\beta_+\rangle$ | $|\beta_-\rangle$ |
| $\beta$ | 2.50 | 2.50 |
| Fidelity | 0.963 | 0.951 |
| $P \times 10^3$ | 19 | 15 |

Table 2. Numerical values of the fidelities and success probabilities taken from Figs. 7 and 9.

|  | 1-1 | | 2-2 | | 4-4 | |
|---|---|---|---|---|---|---|
|  | $|\beta_+\rangle$ | $|\beta_-\rangle$ | $|\beta_+\rangle$ | $|\beta_-\rangle$ | $|\beta_+\rangle$ | $|\beta_-\rangle$ |
| $\beta$ | 1.75 | 1.75 | 2 | 2 | 2.5 | 2.5 |
| Fidelity | 0.963 | 0.977 | 0.981 | 0.958 | 0.984 | 0.98 |
| $P \times 10^3$ | 36 | 56 | 72 | 79 | 35 | 26 |

Table 3. Numerical values of the fidelities and success probabilities taken from Figs. 6 and 8.

## 6. Results

The problem of generating large-size SCSs deserves close attention due to the lack of direct methods for generating them due to the smallness of the nonlinear effects of optical materials while the need for these states is high enough as for testing quantum mechanics and optical QIP. Here we demonstrate efficient ways to implement large-size SCSs in a range from $\beta = 2$ to $\beta = 3$, yet with the highest possible fidelity close to 0.99. Using the $\alpha$-representation of the SCSs allows us to expand the possibilities for engineering of the quantum states. If we know the amplitudes of the SCSs in an arbitrary $\alpha$-representation, we can manipulate and control them as we would do this with standard amplitudes in the Fock basis (0-representation). Instead of generating SCSs, which is a difficult task hardly feasible at the moment, we aim to engineer SCQs being their truncated versions. The optical scheme for generating conditional SCSs in Fig. 1 is fairly simple and requires Fock states as auxiliary ingredients for the quantum state engineering of SCQs. Our numerical results generally show that the larger number of optical elements and auxiliary photons we use, the bigger size and the higher fidelity we can implement the conditional state. In the ideal case with a large set of single photons in auxiliary modes, we can engineer exact SCQs (solid curve in Fig. 3) [21]. But it is unlikely that such an implementation would be acceptable from a practical point of view when a large amount of quantum resources should be consumed. Using Fock states $|n\rangle$ with $n \neq 1$ instead of single photons can well save resources with final results close to ideal (upper bound in Fig. 3). In this sense, the scheme in Fig. 1 with an optical kitten at the entrance can be of more interest since the number of optical BSs, displacement operators and detectors can be reduced to two, while the conditional states still have high qualities (high amplitude and high fidelity). In general, it may be claimed that the scheme in Fig. 1 and corresponding approach can be universally used for the quantum engineering of a wide range of nonclassical states, in particular, for generation of SCSs or amplification of optical kitten. The success probabilities for generation of the desired nonclassical states are commonly small, but it does not matter because the concerned state can be prepared offline until success before their actual use in a certain optical protocol. Moreover, it is quite possible to consider various nonclassical states (e.g., vacuum squeezed states, photon-added coherent states, …) as the inputs not only to the main mode but also to the auxiliary modes. Such intriguing situations deserve a separate investigation.

**Acknowledgement**

S.A.P. is supported by Act 211 Government of the Russian Federation, contract № 02.A03.21.0011, while N.B.A. is supported by the National Foundation for Science and Technology Development (NAFOSTED) under project no. 103.01-2017.08.